\documentclass[11pt]{article}
\usepackage{vietnam,epsfig}

\def\be{\begin{equation}}
\def\ee{\end{equation}}
\def\bea{\begin{eqnarray}}
\def\eea{\end{eqnarray}}

\begin{document}
\vspace*{4cm}
\title{Searches and Electroweak Measurements at HERA}
\author{S. Caron~\footnote{Email: scaron@nikhef.nl}}
\address{NIKHEF, Amsterdam, The Netherlands\\
(on behalf of the H1 and ZEUS collaborations)}

\maketitle\abstracts{
The H1 and ZEUS collaborations have used the HERA I data to
search for physics beyond the SM and to test electroweak physics 
in electron-proton collisions. The new
period of data taking (HERA II) has started and
first HERA II analyses become available.
An overview is given of recent highlights, including
isolated lepton and multi-lepton events, tests of models of new physics
and the first systematic search for deviations
in all final states with high transverse momentum.}

\section{Introduction}
The HERA accelerator at DESY in Hamburg collides 
a $920$~GeV proton beam with a $27.5$~GeV electron beam. 
The resulting
centre-of-mass energy of $319$~GeV and the up to now
recorded integrated luminosity of more than $150 \, \mbox{pb}^{-1}$ is large enough to 
compete with $e^+e^-$ and $p\overline{p}$ 
colliders in testing many aspects of the Standard Model (SM).
The HERA experiments recorded during the HERA I phase (1994-2000)
more than $100 \, \mbox{pb}^{-1}$ each.
Both the HERA collider and experiments have undergone a major upgrade (HERA II
phase) in the years 2000-2001.
New final focusing magnets close to the interaction point 
improve the luminosity and interactions of 
longitudinally polarised electrons or positrons with protons 
can now be studied due to the installation of
spin rotators.
Severe backgrounds prevented the collider experiments from 
efficient data taking up to the end of 2003. 
Since then,
almost $50 \mbox{pb}^{-1}$ integrated luminosity were
recorded by the HERA experiments (till August 2004).
Hence this article presents some updates of HERA I results with 
about $50\%$ more analysed data.

HERA is ideally suited to probe especially electron-quark 
interactions with virtualities of the exchanged boson
in the range of almost $0$ up to around $4 \cdot 10^4~\mbox{GeV}^2$.
The former is called the photoproduction,
the latter the  deep-inelastic scattering domain.
New physics might be produced in both domains, either via resonant 
production of new particles or via virtual effects. 
The HERA data shows sensitivity to various models for new physics. 
In the following I will discuss how well the data agrees
with the SM expectation and present limits 
on a personal selection of models. 

\section{Deep inelastic scattering}
\begin{figure}
\begin{center}
\epsfig{figure=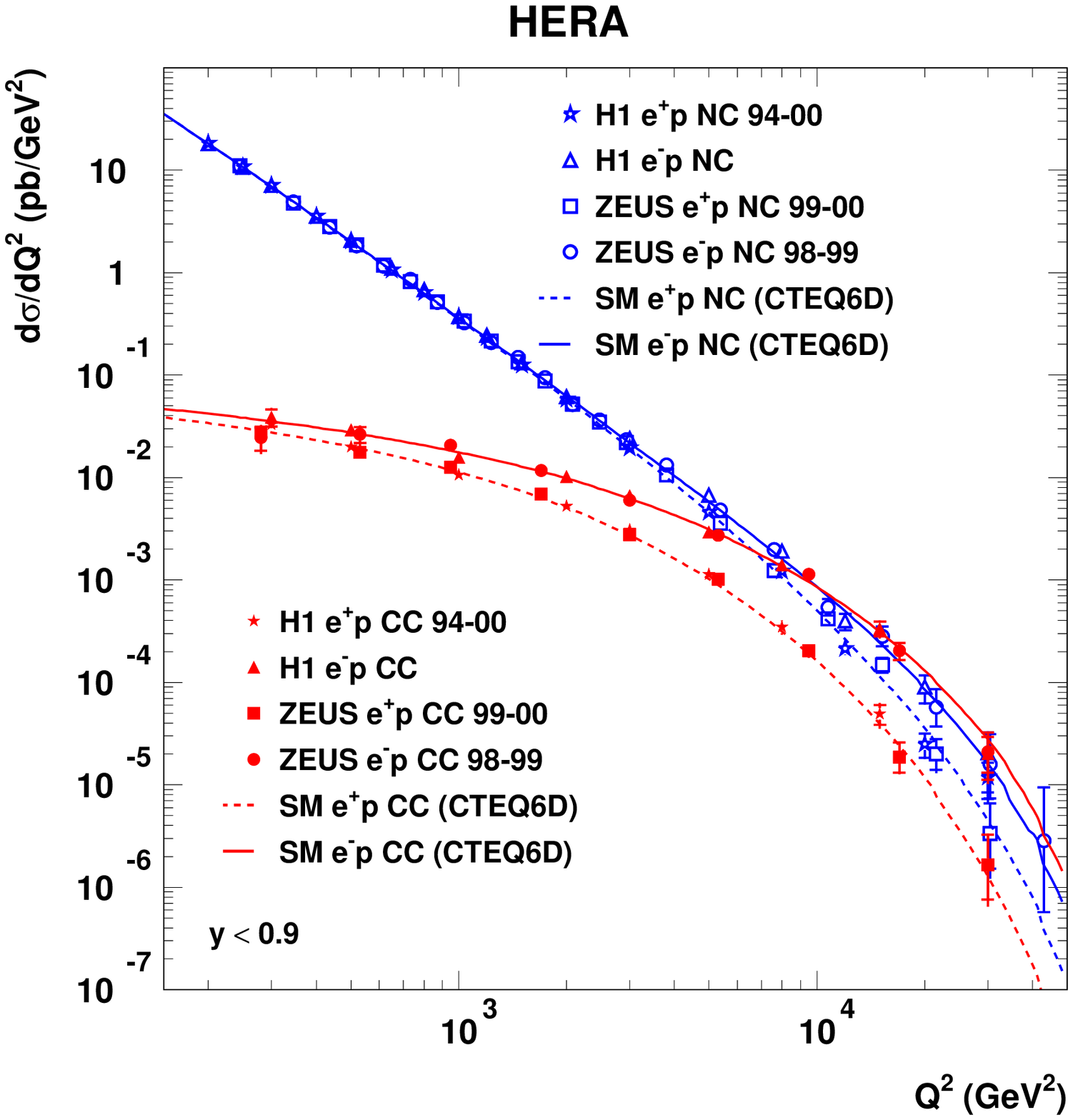,height=8cm}
\epsfig{figure=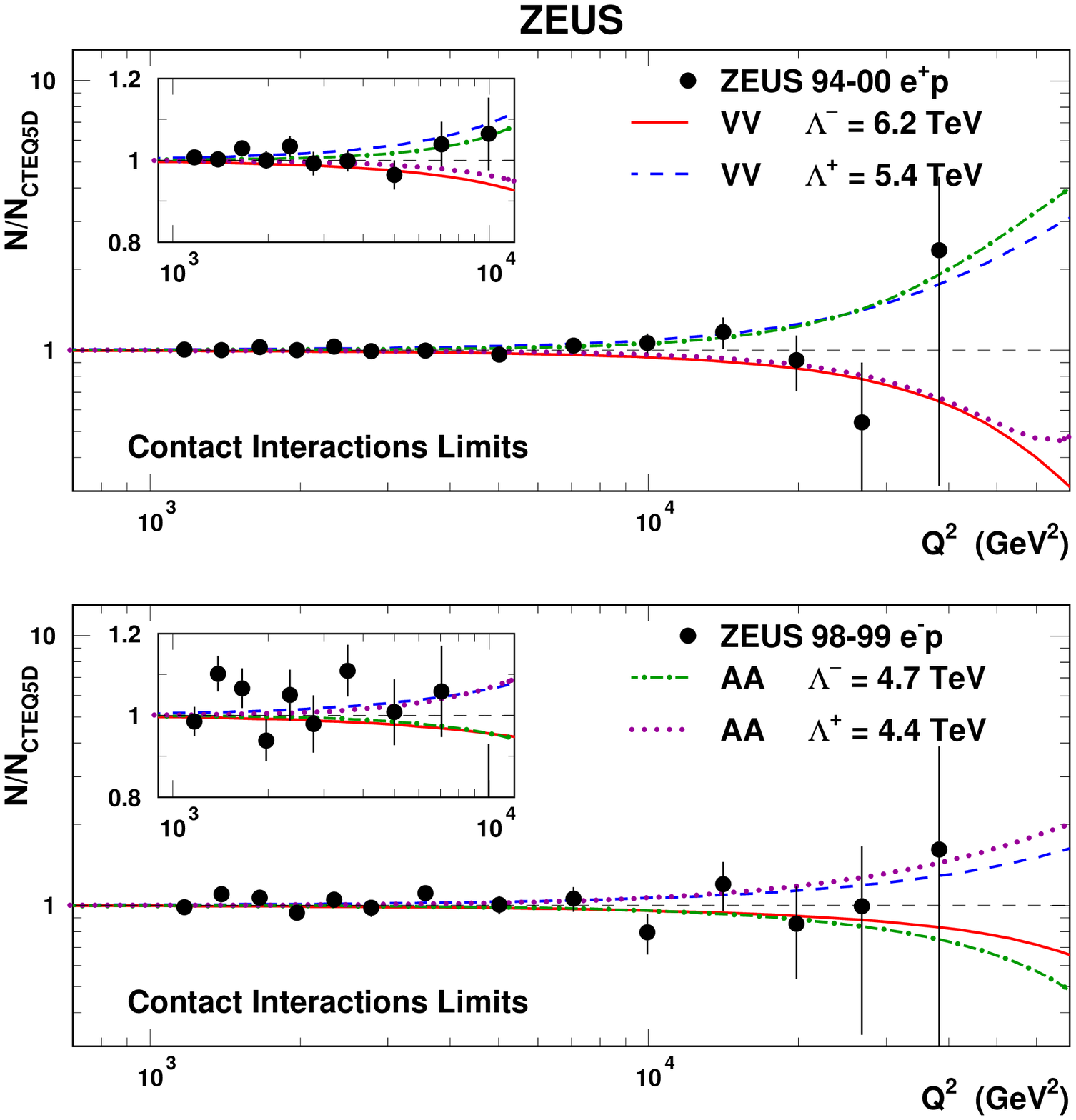,height=8cm}
\end{center}
\caption{Neutral and charged current cross sections as measured by H1 and ZEUS
(left) and a comparison of the ZEUS data to contact interaction models (right).}
\end{figure}
The measured neutral and charged current cross sections are nicely described
by the electroweak SM prediction together with recent parton density functions
and the data of the H1~\cite{Adloff:2003uh} and ZEUS 
~\cite{Chekanov:2003yv} experiments agree well (see Figure 1).
This can be interpreted as a test of the electroweak sector of the SM
with $t-$channel $W$ and photon/$Z^0$ propagators. 
At a virtuality $Q^2\approx M_Z^2$ the charged and neutral
current cross sections are of almost equal size, showing the unification
of weak and electromagnetic forces. 

The data show no significant deviation from the expectation 
and are used to place exclusion limits. 
The $Q^2$ spectrum is typically analysed in the framework 
of so called `contact interactions', where any deviation due to a
new particle or current would correspond to a new heavy mass scale
\cite{Chekanov:2003pw,Adloff:2003jm}. 
The sensitivity of these analyses 
depends on the chiral structure, i.e. the handedness of the $eq$-interaction. 
These general models describe the effects of e.g. 
heavy leptoquarks or lepton compositeness.
Both ZEUS and H1 were able to establish stringent lower limits on the mass
scale $\Lambda$ for positive and negative couplings 
in the range between $1.6$ and $6.2$~TeV. 
The effects of such modifications of the SM are presented in Figure $1$
for different scenarios and for 
positive and negative interference 
with Standard Model processes. A very similar interpretation of these data
leads to lower limits on the mass scale parameters 
of large extra dimension models as proposed by Arkani-Hamed,
Dimopoulos and Dvali of around $0.8$~TeV.
The limit on the radius of the quark charge is around $0.8\times 10^{-18}$~m,
calculated using the classical form factor approximation.

\section{General search for new phenomena}
\begin{figure}
\begin{center}
\epsfig{figure=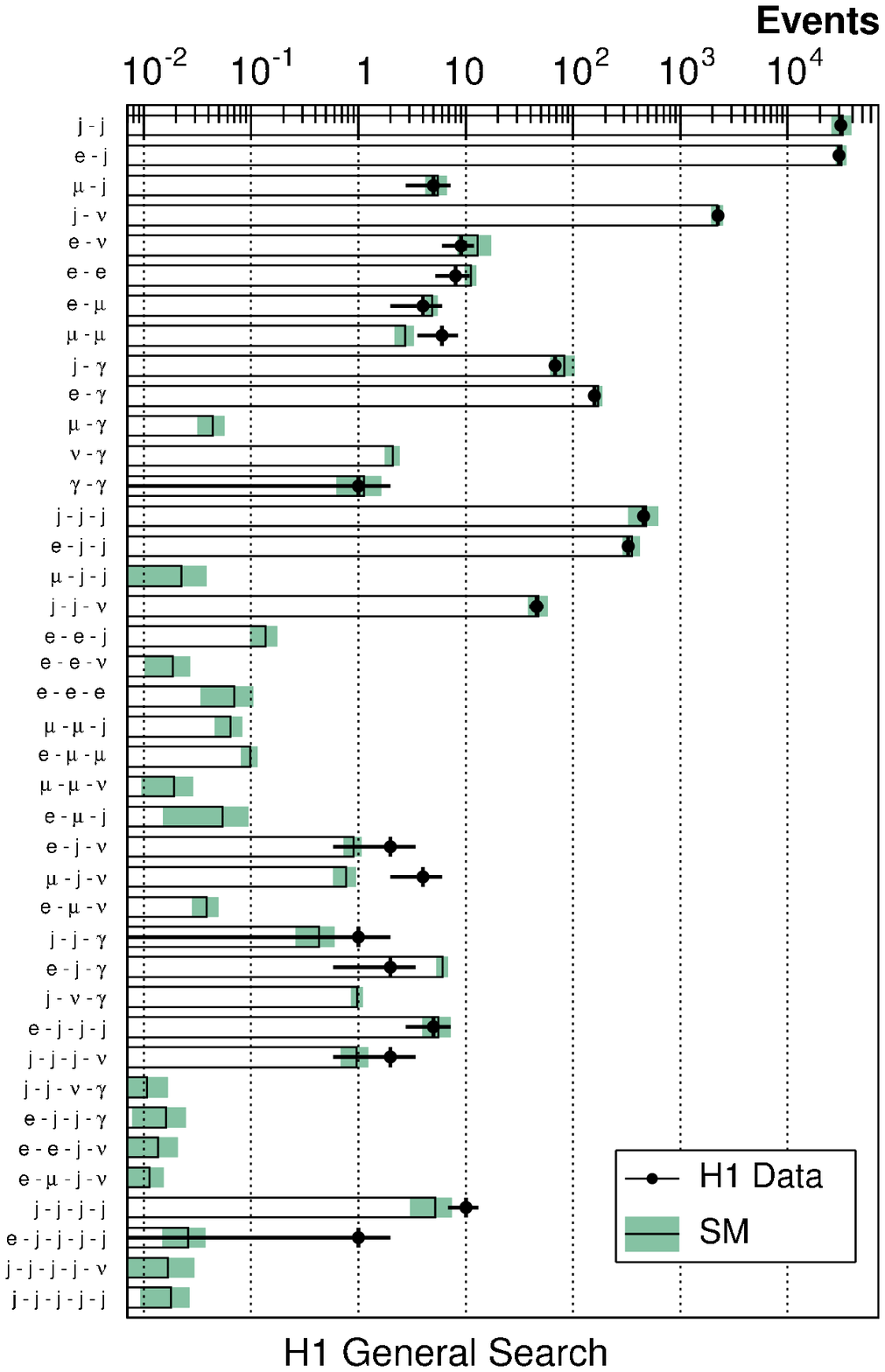,height=9.5cm}
\epsfig{figure=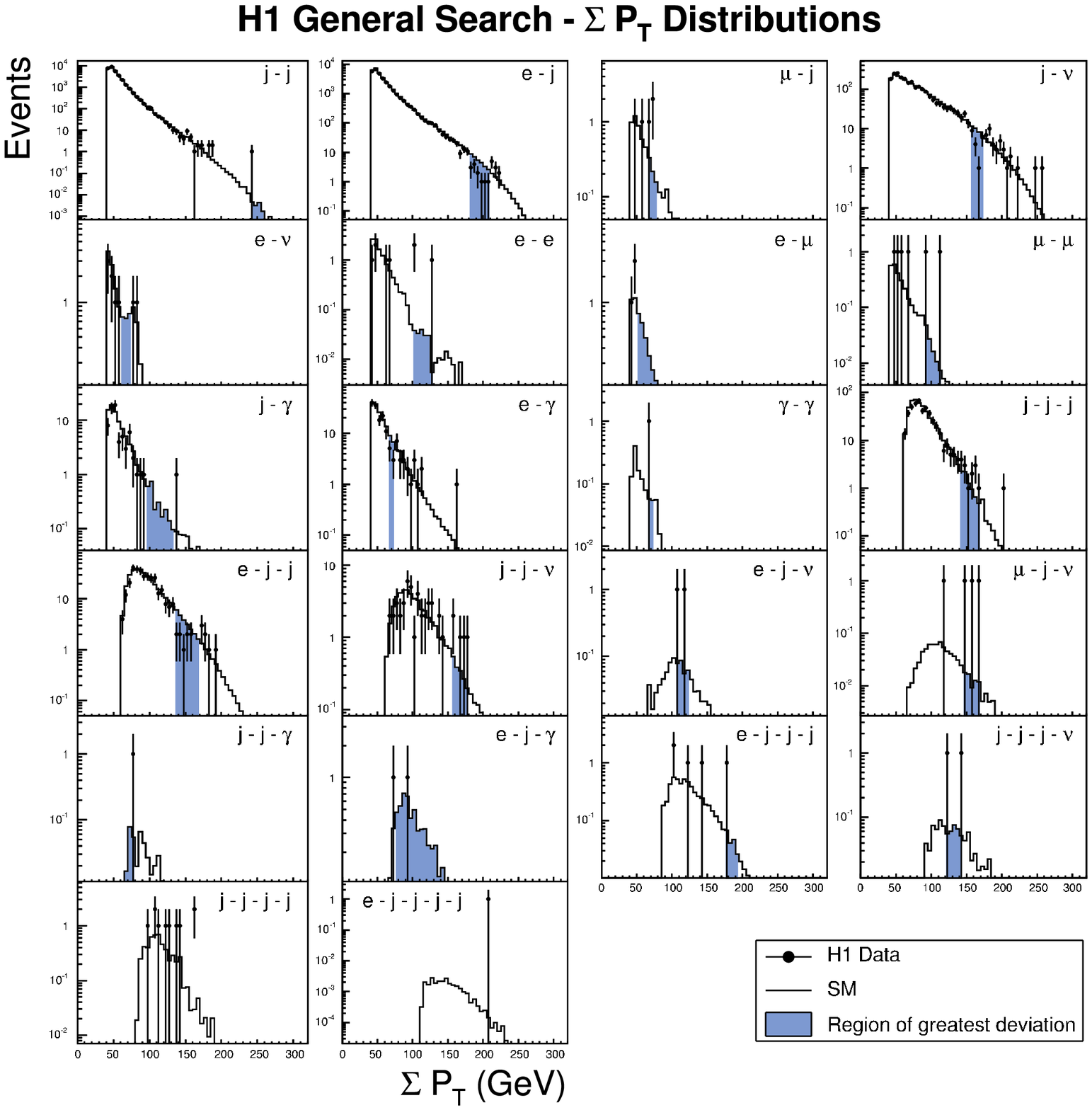,height=9.5cm}
\end{center}
\caption{Comparison of the H1 data measured in various final states with the
SM prediction (left) and the sum of transverse momenta distributions (right).}
\end{figure}
The H1 collaboration has explored all final state topologies with at least
two particles of 20 GeV transverse momentum ($P_T$) in a ``general search''
~\cite{Aktas:2004pz}. 
In this analysis, for the first time, all final state 
configurations accessible to a collider experiment 
are investigated. Such an analysis
could lead to suprises. Many phase space 
corners are not looked at if they are 
not part of a dedicated search
 driven by a specific model.

In order to ensure a clear separation of final states, 
all events are classified into exclusive event classes 
according to the number and types of 
detected particles (e.g. electron-jet or muon-jet-neutrino).
An impressive overall agreement with the predictions is obtained (see figure
$2$), 
illustrating the good understanding of Standard Model physics at HERA 
and of the H1 detector response to the different types of particle. 

In a second analysis step, the distributions of the invariant mass 
and the scalar sum of transverse momenta of the particles are studied 
for each class. Since new physics may be visible as an excess or a 
deficit in some region of one of these distributions, they are 
systematically investigated using a new search algorithm. The algorithm 
locates 
the region with the largest deviation between the data and the SM 
prediction. 
Figure $2$ shows as an example the sum of $P_T$ distributions 
of the data and prediction together with the region selected by the 
algorithm.

To quantify the 
relevance of the disagreement, the probability $\hat{P}$ of observing 
a deviation anywhere in the distribution which is at 
least as significant as that observed is calculated.
No highly significant deviations are observed in most of the event classes. 
The class which shows the largest deviation 
is that with 
a muon, a jet and a neutrino.  
A $\hat{P}$ of below $1\%$ for the invariant mass and about
$0.1\%$ for the sum of $P_T$ distribution is calculated. 
A deviation was also reported previously in this final state in an analysis
especially designed to investigate these kind of events and 
new HERA II results of the analysis are discussed below. 

The final output of the ``general search'' is a list of deviations
and their probability $\hat{P}$. 
In order to get a prediction for these $\hat{P}$ values of the data
the whole analyses are repeated many times with pseudo (Monte Carlo) data.
The analysis of the ``pseudo data'' can give answers to questions
like: ``Given that the SM is correct, in how many H1 experiments
do we expect a deviation of similar of larger size than
the largest deviation observed in all the data distributions?''. 
The answer is that
about $3\%$ ($28\%$ ) of ``pseudo data'' sets 
whould have a larger deviation 
in the sum of $P_T$ (invariant mass)
distributions. 

It is highly interesting to see how these values behave 
for the upcoming HERA II data.

\section{Isolated lepton events and single top production}
A couple of interesting events have been observed in HERA I data in the
analysis of events with a lepton, a jet and missing transverse momentum
\cite{Andreev:2003pm}. 
More data events than expected were seen by the H1 experiment for 
$P_{T}^X>25~$GeV, where $P_{T}^X$ is the transverse momentum of the 
hadronic system. 
The analysis has been updated with recent HERA II data~\cite{hera2}. H1 finds 
in $163 \mbox{pb}^{-1}$ integrated luminosity
$14$ events with an electron or muon,
while $5.1\pm 1.0$ are expected.
The SM prediction for these events is dominantly the  
production of $W$ bosons with high $P_T$. In the HERA I ZEUS data,
in total $7$ events are found in the electron and muon channels, while
$5.6 \pm 0.7$ are expected~\cite{Chekanov:2003yt}.
A clarification of these differences will require a larger data set.

These data have triggered model comparisons, e.g. a 
flavour-changing neutral-current production of top 
quarks at HERA and the subsequent 
leptonic W decay could 
lead to the event signature of the isolated lepton events.
Top quark production via
anomalous couplings between the u-quark, a photon and the top ($\kappa_{tu\gamma}$) or between the u-quark, the $Z$ and the top ($v_{tuZ}$) have been studied
by H1 and ZEUS using optimized cuts~\cite{Aktas:2003yd,Chekanov:2003yt}. 
The H1 analysis finds again the excess in the electron and muon channels 
and no deviation in the jet channel. The ZEUS analysis finds 
agreement between data and expectation in all three channels.
Stringent limits on anomalous top couplings have been derived as
shown in Figure $3$.

\begin{figure}
\begin{center}
\epsfig{figure=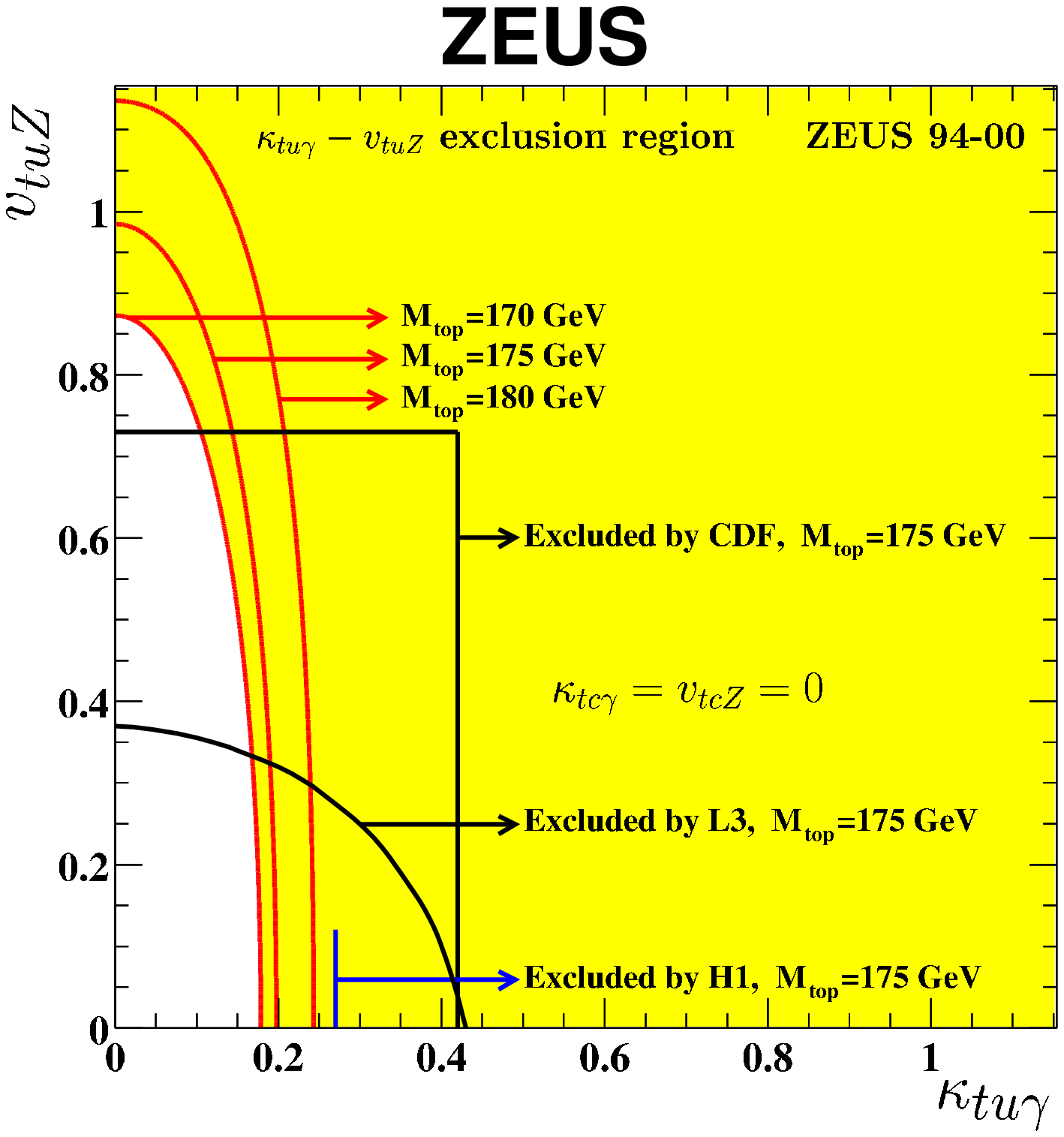,height=7cm}
\end{center}
\caption{Exclusion regions at $95\%$ confidence level for anomalous top
couplings.}
\end{figure}

\section{Multi-lepton events}
H1 reports furthermore $3$ interesting
di-electron events with an invariant mass
of the di-electron system $M_{ee}>100$~GeV in $163\mbox{pb}^{-1}$ HERA I and
HERA II data~\cite{hera2}. The SM expectation was
found to be $0.44\pm 0.1$. In the HERA I data recorded by ZEUS ($130\mbox{pb}^{-1}$) also 2 events are found with  $M_{ee}>100$~GeV, while 
$0.77 \pm 0.08$ are expected ~\cite{zeuse}. No deviation is found 
in the di-muon data of both experiments.
In the same H1 dataset $3$ tri-electron events with an invariant
mass of the two highest $P_T$ electrons $M_{12}>100$~GeV are found.
The expectation is $0.31\pm 0.08$. 
In the HERA I data the ZEUS experiment  
expects $0.37\pm 0.04$ and finds no tri-electron event with $M_{12}>100$~GeV.
Again conclusions on these events 
can only be drawn with a significant amount
of additional data.


\section{Summary}
Recent HERA search highlights are presented, including updates using HERA II 
data. The HERA I data set has been exploited using
searches for deviations in all final state configurations. Although some
interesting events have been observed
no evidence for new physics could be established so far. 
Some tests of specific models are presented as examples and underline that
HERA data can still give useful information in testing the SM and probing
the physics which might be beyond.

HERA data has also been used to determine exclusion limits~\cite{Kuze:2002vb}
 on 
e.g. $R_P$ violating 
Supersymmetry, Leptoquarks, 
doubly charged Higgs production, excited fermions and magnetic monopoles
and details can be found in recent H1 and ZEUS publications.

\section*{Acknowledgments}
The author would like to thank the organisers for a very enjoyable conference.
The author acknowledges the support of a
 Marie Curie Intra-European Fellowship 
within the 6th European Community Framework Programme.
The author thanks  
Elisabetta Gallo, Emanuelle Perez and  Alex Tapper for comments on the manuscript.

\end{document}